\documentclass[twocolumn]{aastex62}

\usepackage{times}
\usepackage{amsmath}
\usepackage{graphicx}
\usepackage{hyperref}
\usepackage{gensymb}
\usepackage{upgreek}
\usepackage{natbib}
\usepackage{subfigure}
\usepackage{multirow}
\usepackage{comment}
\usepackage{mathtools}
\usepackage{mathrsfs}
\usepackage{fontenc}
\usepackage{color}
\usepackage{url}
\usepackage{pifont}

\newcommand{\HII}{H\,\textsc{ii}}
\newcommand{\NeII}{[Ne\,\textsc{ii}]}
\newcommand{\NeIII}{[Ne\,\textsc{iii}]}
\newcommand{\NeV}{[Ne\,\textsc{v}]}
\newcommand{\OII}{[O\,\textsc{ii}]}
\newcommand{\OIII}{[O\,\textsc{iii}]}

\newcommand{\Lbol}{$L_{\rm bol}$}
\newcommand{\SII}{[S\,\textsc{ii}]}
\newcommand{\NII}{[N\,\textsc{ii}]}

\submitjournal{ApJ}

\shorttitle{SFR in Active Galaxies}
\shortauthors{Zhuang, Ho \& Shangguan}

\begin{document}

\title{A New Method to Measure Star Formation Rates in Active Galaxies Using 
Mid-infrared Neon Emission Lines}

\author[0000-0001-5105-2837]{Ming-Yang Zhuang}
\email{mingyangzhuang@pku.edu.cn}
\affil{Kavli Institute for Astronomy and Astrophysics, Peking University,
Beijing 100871, China}
\affil{Department of Astronomy, School of Physics, Peking University,
Beijing 100871, China}

\author[0000-0001-6947-5846]{Luis C. Ho}
\affil{Kavli Institute for Astronomy and Astrophysics, Peking University,
Beijing 100871, China}
\affil{Department of Astronomy, School of Physics, Peking University,
Beijing 100871, China}

\author[0000-0002-4569-9009]{Jinyi Shangguan}
\affil{Max-Planck-Institut f\"{u}r extraterrestrische Physik, 
Gie{\ss}enbachstr. 1, D-85748 Garching, Germany}

\begin{abstract}

The star formation rate (SFR) is one of the most fundamental parameters of 
galaxies, but nearly all of the standard SFR diagnostics are difficult to 
measure in active galaxies because of contamination from the active galactic 
nucleus (AGN). Being less sensitive to dust extinction, the mid-infrared 
fine-structure lines of \NeII\ 12.81 \micron\ and \NeIII\ 15.56 \micron\ 
effectively trace the SFR in star-forming galaxies.  These lines also have the 
potential to serve as a reliable SFR indicator in active galaxies, provided 
that their contribution from the AGN narrow-line region can be removed.  We 
use a new set of photoionization calculations with realistic AGN spectral 
energy distributions and input assumptions to constrain the magnitude of 
\NeII\ and \NeIII\ produced by the narrow-line region for a given strength of 
\NeV\ 14.32 \micron.  We demonstrate that AGNs emit a relatively restricted 
range of \NeII/\NeV\ and \NeIII/\NeV\ ratios.  Hence, once \NeV\ is measured, 
the AGN contribution to the low-ionization Ne lines can be estimated, and the 
SFR can be determined from the strength of \NeII\ and \NeIII. We find that AGN 
host galaxies have similar properties as compact extragalactic \HII\ regions, 
which indicates that the star formation in AGN hosts is spatially concentrated.
This suggests a close relationship between black hole accretion and nuclear 
star formation.  We update the calibration of \NeII\ and \NeIII\ strength as a 
SFR indicator, explicitly considering the effects of metallicity, finding very 
good relations between Ne fractional abundances and the \NeIII/\NeII\ ratio 
for different metallicities, ionization parameters, and starburst ages.  
Comparison of neon-based SFRs with independent SFRs for active and star-forming
galaxies shows excellent consistency with small scatter ($\sim0.18$ dex). 
\end{abstract}


\section{Introduction} \label{sec:intro}

The empirical correlations between the masses of central black holes and the 
properties of their host galaxies \citep[e.g.,][]{2000ApJ...539L..13G, 
2000ApJ...539L...9F, 2013ARA&A..51..511K} have led to the widely accepted 
idea that the growth of central black holes must be connected to galaxy evolution 
\citep[e.g.,][]{1998Natur.395A..14R, 2004coa..book.....H, 2005Natur.433..604D, 
2006MNRAS.370..645B, 2011MNRAS.410...53F, 2014ARA&A..52..589H}.  
Discussions of the coevolution of active galactic nuclei (AGNs) and their hosts 
usually takes two forms: (1) the weak form: galaxies affect the growth of black 
holes by controlling AGN accretion and merging via global, galaxy-wide 
processes; (2) the strong form: black holes control galaxy properties via energy 
and momentum feedback into their large-scale environment \citep
{2012NewAR..56...93A, 2012ARA&A..50..455F, 2013ARA&A..51..511K}. 
Studying the growth of black holes and their host galaxies is the key to understanding 
this coevolution.  From an observational point of view, one of the major challenges 
is to measure the star formation rate (SFR) of AGN host galaxies.

A number of methods to obtain precise SFRs have been developed for 
star-forming galaxies, including diagnostics involving the ultraviolet (UV) 
and infrared (IR) continuum, as well as emission-line strengths that trace 
the budget of ionizing photons from massive stars (see \citealt [for a 
review]{1998ARA&A..36..189K}). Unfortunately, none of these methods can be 
applied unambiguously to active galaxies.  AGNs emit copious ionizing and 
high-energy photons across its broad-band spectral energy distribution (SED), 
which greatly affect the ionization balance of the line-emitting gas exposed 
to it. This compromises most of the traditional SFR diagnostics, if not 
rendering them essentially useless.  For instance, photoionization by a central
AGN source produces strong hydrogen recombination lines (e.g., H$\alpha$) and 
forbidden lines (e.g., \OII\ $\lambda$3727). In extreme cases, the narrow-line 
region (NLR) of the AGN can occupy the entire extent of the host galaxy 
\citep{2011ApJ...732....9G}.  Dust grains, on circumnuclear (torus) scales 
and beyond, absorb and reprocess high-energy photons from the nucleus, 
generating thermal radiation across a broad IR spectrum.

Neon shares a nucleosynthetic history similar to that of oxygen, and 
is the third most abundant metal in the interstellar medium next to oxygen 
and carbon. With an ionization potential of 21.56 eV, \NeII\ 12.81 \micron\ 
acts as one of the primary coolants in \HII\ regions.  In environments of 
lower metallicity, \NeIII\ 15.56 \micron,
which has an ionization potential of 40.96 eV, becomes more dominant.  These 
mid-IR neon lines suffer significantly less extinction than conventional UV 
and optical transitions.  \citet{2007ApJ...658..314H} show that the sum of 
\NeII\ and \NeIII\ correlates tightly with the far-IR luminosity and Br$\alpha$
luminosity in \HII\ regions and star-forming galaxies, and hence can be used as
an excellent tracer of star formation.  However, as the NLR of AGNs can also 
produce \NeII\ and \NeIII, these low-ionization lines alone cannot serve as 
robust indicators of the SFR in active galaxies.   Fortunately, the ionization 
potential of 97.12 eV for \NeV\ 14.32 \micron, which lies sandwiched between 
\NeII\ and \NeIII, is sufficiently high that, under almost all circumstances, 
it can be uniquely attributed to AGN photoionization alone \citep[e.g.,][]{
2007ApJ...656..148A}.  The only exception might be environments in which fast 
shocks contribute to the excitation of the gas \citep[e.g.,][]{2002A&A...383...46M, 
2011ApJ...726...86R}.  Hence, so long as 
\NeV\ is visible, we can be reasonably confident that an AGN is 
present\footnote{The converse is not true.  The absence of \NeV\ does not mean 
that there is no AGN, for the AGN can be of low ionization (e.g., low-ionization 
nuclear emission-line regions), as typifies systems with low mass accretion 
rates \citep{2008ARA&A..46..475H, 2009ApJ...699..626H}}.

Whenever all three mid-IR neon lines are detected, can we estimate the amount 
of \NeII\ and \NeIII\ emission expected from AGN photoionization, given the 
observed strength of \NeV?  As we know that \NeV\ can only be produced by 
AGNs, if we can reliably use theoretical calculations to bracket the level of 
the lower-ionization transitions associated with \NeV, we can ascertain whether
there is any ``excess'' \NeII\ and \NeIII\ emission beyond that expected 
from the NLR, which then can be attributed to star formation.  This strategy 
closely follows that of \citet{2005ApJ...629..680H}, who proposed a method of 
estimating SFRs in AGNs based on \OII\ $\lambda$3727, using \OIII\ 
$\lambda$5007 to predict the relative strength of \OII\ $\lambda$3727 expected 
from photoionization in high-ionization AGNs \citep{2006ApJ...642..702K}.

Numerous attempts have been made to analyze the narrow-line spectrum of AGNs in
the context of photoionization models \citep[e.g.,][]{1986ApJ...300..658F, 
1993ApJ...410..567H, 2004ApJS..153....9G, 2006A&A...458..405G, 
2014MNRAS.438..901S}. 
This paper focuses on a limited goal: to use realistic AGN photoionization models to 
predict the maximum range of the strengths of \NeII\ 12.81 \micron\ and \NeIII\
 15.56 \micron\ relative to \NeV\ 14.32 \micron.   We show that 
in the radiation pressure-dominated regime, AGNs produce a 
relatively stable, restricted range of mid-IR neon emission-line spectra,
paving the way for an updated calibration of \NeII\ and \NeIII\ as a new SFR 
indicator for active galaxies.  Our neon-based SFRs show good consistency with 
independent SFRs based on SED fitting.

\section{Photoionization Model}  \label{sec:model}

We perform photoionization calculations using the latest version (C17.01) of 
{\tt CLOUDY} \citep[][]{2017RMxAA..53..385F}.  {\tt CLOUDY} is designed to 
predict the observed spectrum of interstellar clouds by solving the equations of 
statistical and thermal equilibrium to predict their thermal, ionization, and chemical 
structure for a given set of input conditions.

\begin{figure}[t]
\centering
\includegraphics[width=0.48\textwidth]{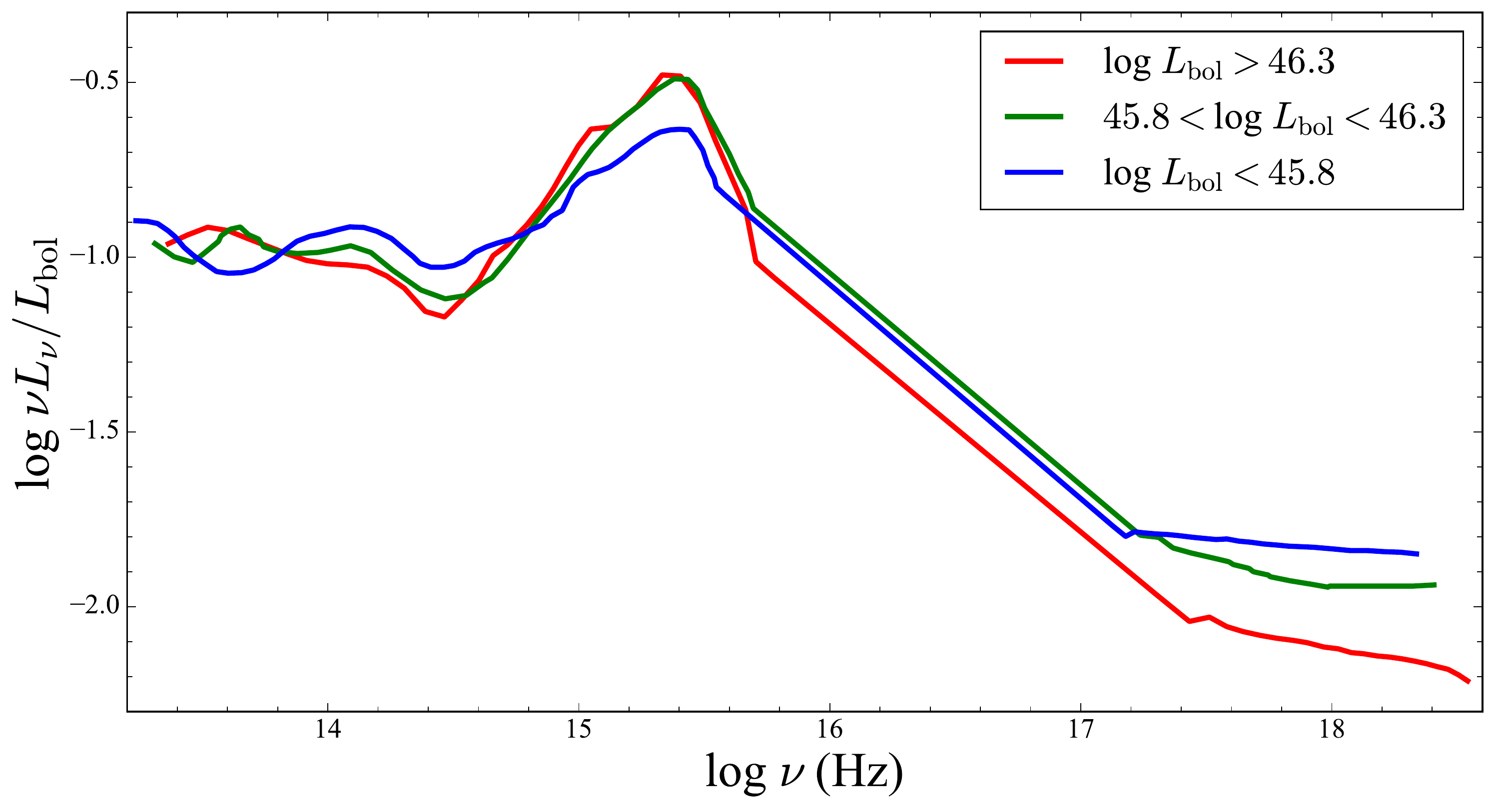}
\caption{Median input AGN SED for log (\Lbol/erg s$^{-1}$) $>$ 46.3 (red), 
45.8$-$46.3 (green), and $<$ 45.8 (blue).  Adapted from 
\citet{2014MNRAS.438.2253S}.}
\label{fig1}
\end{figure}

\subsection{Incident Radiation Field}  \label{subsec:spe-shape}

Early photoionization models of AGNs typically assume a canonical power-law 
ionizing spectrum ($F_{\nu} \propto \nu^{\alpha}$) with an IR and high-energy 
cutoff \citep[e.g.,][]{1986ApJ...300..658F, 2004ApJS..153....9G}. However, this
estimation does not fully describe all the key features of the SED, including 
the ``big blue bump,'' whose peak and detailed shape in the far-UV and 
extreme-UV affect the line ratios studied here.  To properly constrain the big 
blue bump, one must construct a composite SED by combining multi-wavelength 
data of objects covering a large range of redshift to ensure coverage of 
the rest-frame region shortward of 912 \AA.
\citet{2014MNRAS.438.2253S} assembled a large sample of 761 type 1 AGNs 
spanning a range of redshifts from 0.11 to 3.29, with data covering up to 19 
photometric bands.  Studying a subset of 231 objects with high-quality SEDs,
they found that the peak of the big blue bump shifts slightly but systematically
toward longer wavelengths with increasing redshift. Because luminosity 
correlates with redshift in this flux-limited sample, the systematic dependence
of the SED on redshift translates into a dependence on luminosity.  The peak of
the big blue bump for low-$z$, lower luminosity objects lies at $\sim$1100 \AA,
similar to that reported in \citet{2005ApJ...619...41S}, who analyzed 17 
AGNs with quasi-simultaneous spectrophotometry covering rest-frame $900-9000$ 
\AA.  With increasing redshift or luminosity, the contribution of the UV to 
the bolometric luminosity increases, while X-rays become less important.  This
overall trend is consistent with simultaneous UV, optical, and X-ray 
observations that find that the relative flux between X-rays and UV decreases 
with increasing UV luminosity \citep[e.g.,][]{2012ApJS..201...10W}.  As input 
incident radiation fields for our calculations, we choose three median SEDs 
from \citet{2014MNRAS.438.2253S} binned by bolometric luminosity: 
log (\Lbol/erg s$^{-1}$) $>$ 46.3, 45.8$-$46.3, and $<$ 45.8 (Figure \ref{fig1}).  
With increasing \Lbol, the peak of the big blue bump shifts toward longer
wavelength, and a greater fraction of the energy emerges in the UV relative to
the X-rays.

\subsection{Narrow-line Region Properties}

We assume isobaric conditions with radiation pressure and thermal pressure 
mainly contributing to the total pressure.  The NLR clouds have a 
plane-parallel geometry and coexist with dust grains \citep[e.g.,][]{
2003ApJ...587..117R}.  We adopt the default solar metallicity abundances 
and dust depletion factors from {\tt CLOUDY C17.01}\footnote{Values in 
Tables 7.1 and 7.8 of the {\tt CLOUDY} documentation Hazy 1.}, except for 
the abundance of nitrogen, which is scaled to twice the solar value, following 
\citet{2006ApJ...642..702K}. 
The composition and size distribution of the dust grains are similar to those 
of Orion. The size-resolved polycyclic aromatic hydrocarbons (PAHs) have the 
same size distribution as in \citet{2008ApJ...686.1125A}.  We consider 
photoelectric heating of the gas by dust, which is very important in determining 
the temperature structure of the NLR \citep{2002ApJ...572..753D}.  The ionization 
parameter, defined as the dimensionless ratio of the incident ionizing photon 
density to the hydrogen density, $U \equiv {\Phi_{\rm H}}/{n_{\rm H} c}$, 
where $\Phi_{\rm H}$ is the flux of hydrogen ionizing photons per second, 
$n_{\rm H}$ is the hydrogen density, and $c$ is the speed of light. We consider
two initial values for the hydrogen density, $n_{\rm H} = 10^2$ and $10^3$ 
cm$^{-3}$, which covers most of the values inferred for Seyfert galaxies based
on the \SII\ $\lambda6716/\lambda6731$ ratio \citep{2013MNRAS.430.2605Z}.  
As we are interested in powerful AGNs with detected \NeV\ emission, sources 
with relatively high ionization parameter, we restrict our investigation to 
log $U \approx -2.5$ to $-0.1$, in steps of 0.3 dex.  This range of ionization 
parameters is typical of the NLR in classical Seyfert nuclei \citep[e.g.,][]
{1993ApJ...410..567H} and low-redshift type 2 quasars 
\citep{2008MNRAS.390..218V}.

\section{Results and Discussions} \label{sec:results}

\begin{figure}[t]
\centering
\includegraphics[width=0.48\textwidth]{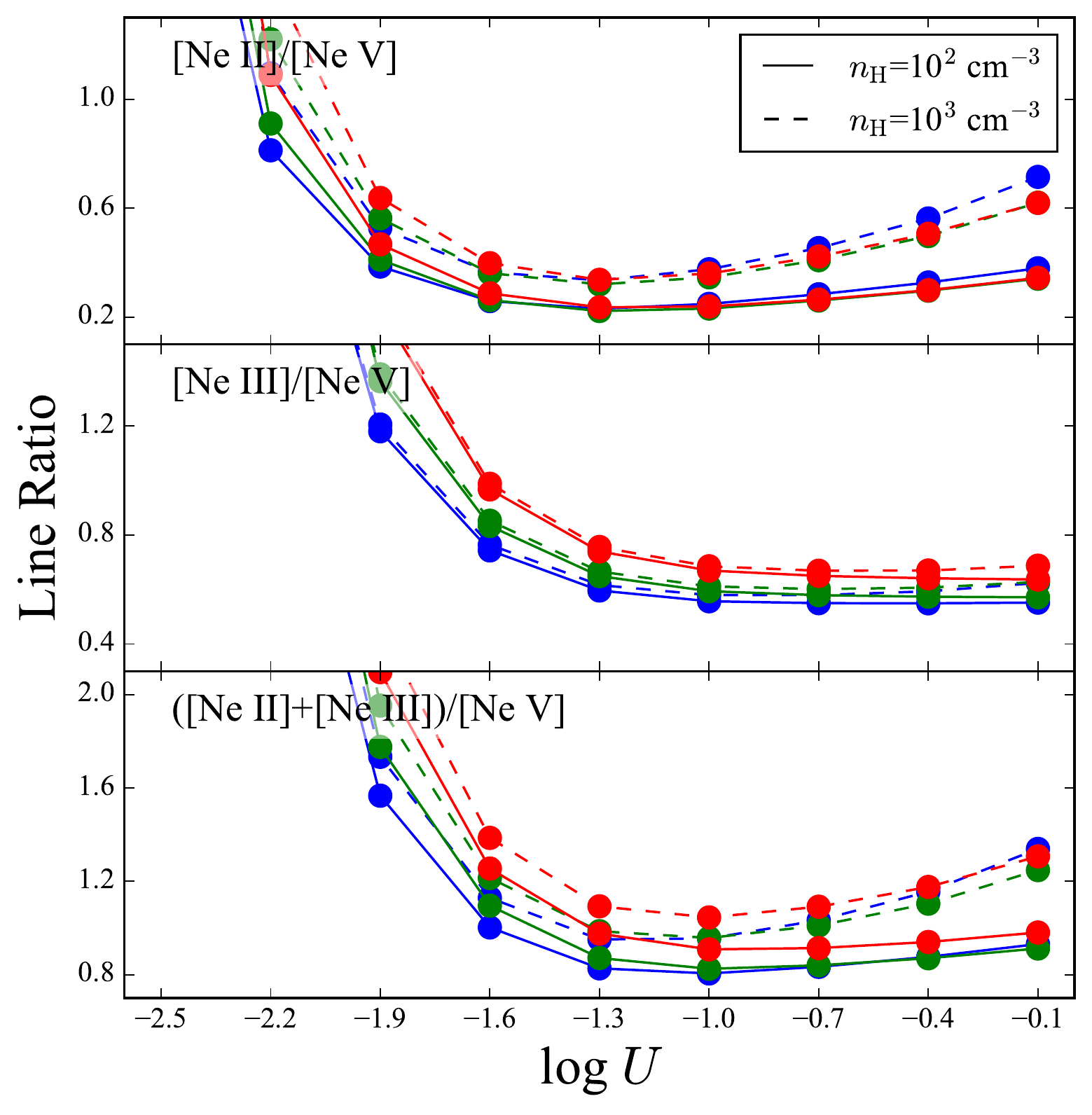}
\caption{The dependence of (top) \NeII/\NeV, (middle) \NeIII/\NeV, and 
(bottom) (\NeII+\NeIII)/\NeV\ on ionization parameter $U$, for hydrogen particle
densities $n_{\rm H}=10^2$ cm$^{-3}$ (solid lines) and $n_{\rm H} = 10^3$ 
cm$^{-3}$ (dashed lines), and the three input SEDs shown in Figure \ref{fig1}: 
log (\Lbol/erg s$^{-1}$) (red) $>$ 46.3, (green) 45.8$-$46.3, and (blue) 
$<$ 45.8.}
\label{fig2}
\end{figure}

\subsection{AGN Contribution to Mid-IR Neon Lines} \label{agn_contribution}

Figure \ref{fig2} shows the influence of the ionization parameter on the 
strengths of the low-ionization lines \NeII\ and \NeIII\ relative
to the high-ionization line \NeV.  
Remarkably, all the line ratios remain stable when log $U \gtrsim -1.6$, 
with a scatter of $\lesssim 0.16$, within the range of hydrogen densities and 
ionizing spectra considered in our model grid. This 
behavior is consistent with that of the dusty, radiation pressure-dominated 
isobaric photoionization models of \citet{2002ApJ...572..753D}, in which the 
ionization parameter-sensitive line ratio \OIII\ $\lambda5007$/H$\beta$ shows 
only mild variation with $U$ in the regime of high $U$.
When $U \gtrsim 0.03$ \citep{2014MNRAS.438..901S}, 
dust becomes the dominant source of opacity and 
effectively couples radiation pressure to gas pressure.  Dust influences the 
temperature structure of the NLR via photoelectric heating and excitation of 
the high-ionization gas.  Radiation becomes the dominant source of pressure
contributing to the static gas pressure, and the gas density scales linearly 
with ionizing flux, making the local ionization parameter independent of the 
initial ionization parameter \citep{2002ApJ...572..753D}.  
In any event, the line ratios depicted in Figure \ref{fig2} are 
affected only mildly by the range of densities explored in our calculations.  
As shown in \citet{2002ApJ...572..753D} and 
\citet{2014MNRAS.438..901S}, the small dispersion of derived $U$ for 
Seyferts suggests that the NLRs of most AGNs are radiation pressure-dominated.
In the radiation pressure-dominated regime ($-1.6 \leq$ log $U$ $\leq -0.1$) 
and over the range of parameters investigated here, we find

\begin{equation} \label{eq1}
{\rm \NeII/\NeV = 0.345 \pm 0.118}
\end{equation}
\begin{equation} \label{eq2}
{\rm \NeIII/\NeV = 0.628 \pm 0.109}
\end{equation}
\begin{equation} \label{eq3}
{\rm (\NeII+\NeIII)/\NeV = 0.987 \pm 0.157}.
\end{equation}

\noindent
This indicates that AGNs that span typical NLR conditions (in terms of ionizing 
SED, ionization parameter, and density) emit a relatively restricted range of 
\NeII\ and \NeIII\ for a given level of \NeV.  
Similar trends are found by \citet {2006A&A...458..405G}
and \citet{2014MNRAS.438..901S}. The absolute value of their 
line ratios differs somewhat from ours mainly because of the choice of different
input ionizing spectra. As described in Section \ref{subsec:spe-shape}, our 
adopted incident radiation field is based on observed intrinsic SEDs of 
unobscured AGNs that span a wide range in bolometric luminosity.

The above results are welcomed news from the point of view of our primary 
science goal of using the low-ionization neon lines to estimate the SFR in AGN 
host galaxies.  A stable \NeII/\NeV\ and \NeIII/\NeV\ ratio from the NLR 
ensures that we can use the detected \NeV\ emission to forecast---and thereby 
subtract---the amount of \NeII\ and \NeIII\ emission originating from the AGN, 
leaving the remainder, if any, to star formation.  We stress, once again, that 
this method breaks down for low-ionization sources,
which, by definition, exhibit weak high-ionization species compared to 
low-ionization transitions.

\subsection{Star Formation in AGN Host Galaxies} \label{AGN hosts}

In order to test the reliability of our results, we use the galaxy sample of 
\citet{2012ApJ...758....1L}, which contains 33 AGNs with \NeII, \NeIII, and 
\NeV\ detections and 79 star-forming galaxies with both \NeII\ and \NeIII\ 
detections obtained from high-resolution spectra observed with the 
{\it Spitzer}/Infrared Spectrograph (IRS).  We also augment this sample with 
103 nuclear and extranuclear high-resolution IRS spectra from the {\it Spitzer}\
Infrared Nearby Galaxies Survey \citep[SINGS;][]{2009ApJ...693.1821D}, after 
excluding objects classified as AGNs and those lacking \NeII\ or \NeIII\ detections.

\begin{figure}[t]
\centering
\includegraphics[width=0.48\textwidth]{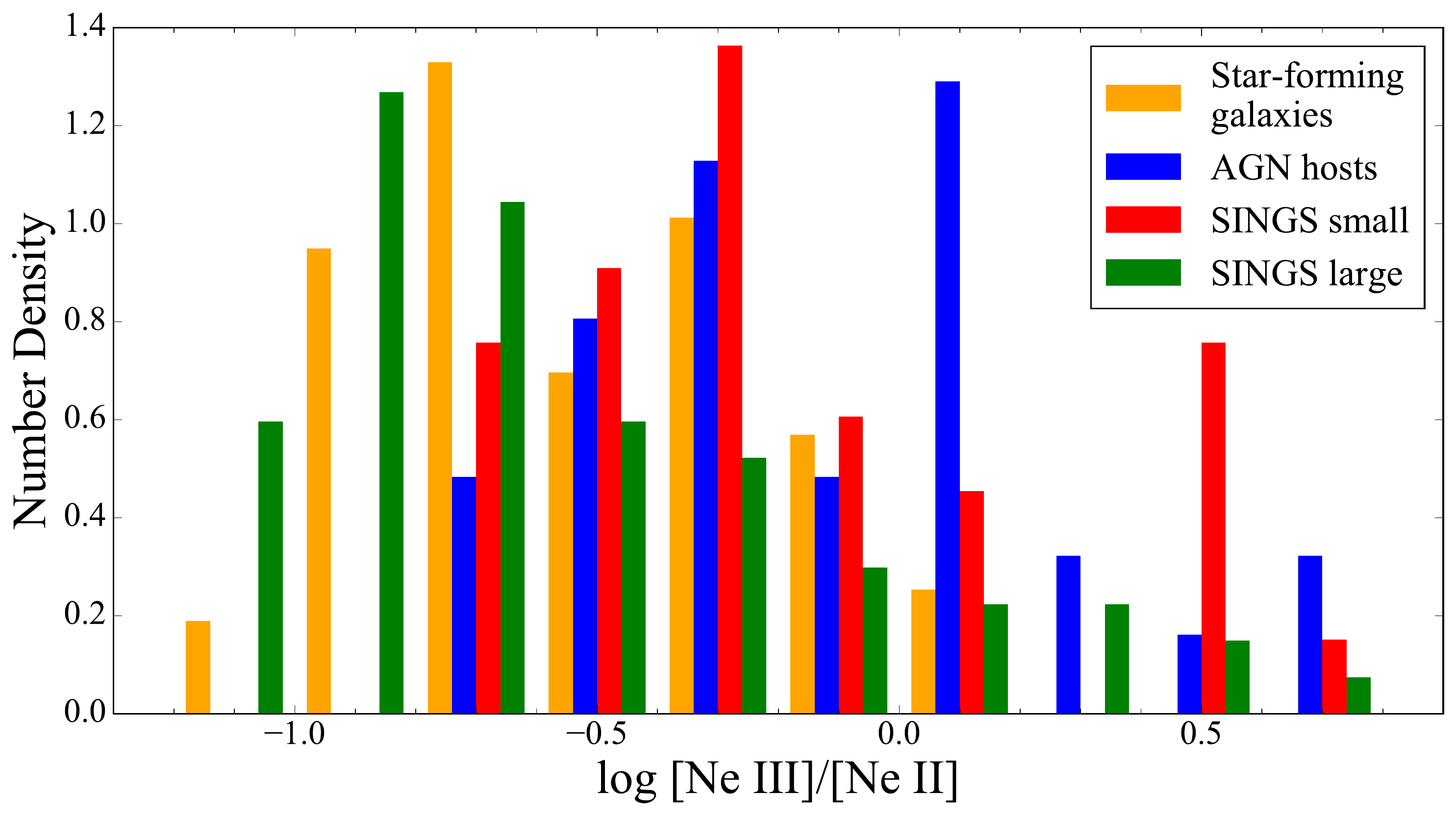}
\caption{Distribution of \NeIII/\NeII\ ratios for star-forming galaxies 
(orange), AGN host galaxies (blue), and SINGS emission-line regions with 
projected extraction area $<$ 0.3 kpc$^2$ (SINGS small; red) and $\geq$ 0.3 
kpc$^2$ (SINGS large; green), normalized by the total number of objects in 
each subsample. The measurements for the AGN hosts have been corrected for 
AGN contamination.}
\label{fig3}
\end{figure}

Figure \ref{fig3} shows the distribution of \NeIII/\NeII\ for star-forming 
galaxies, AGN hosts, and emission-line regions extracted from SINGS. The ratios 
for the AGN host galaxies were obtained after subtracting the NLR contribution 
of the AGN using Equations \ref{eq1} and \ref{eq2}. The SINGS sources are 
further divided into two subgroups: 34 targets with spectra extracted over a 
relatively small projected area of $< 0.3$ kpc$^2$ (SINGS small), and 69 
targets with a larger extraction area of $\geq$ 0.3 kpc$^2$ (SINGS large).  For
star-forming galaxies, \NeIII/\NeII\ spans $\sim 0.1 - 1$, with a median 
value of 0.26.  AGN hosts have a higher median of \NeIII/\NeII\ = 0.66, but 
their distribution still largely overlaps with that of star-forming galaxies.
Interestingly, the two subgroups of SINGS spectra show different distributions:
objects extracted from a larger physical area tend to have \NeIII/\NeII\ ratios
similar to those of star-forming galaxies, while those extracted from smaller 
areas closely resemble AGN hosts.  A Kolmogorov-Smirnov test of the 
distributions of \NeIII/\NeII\ for AGN hosts and the ``SINGS small" sample
indicates that the null hypothesis that the two distributions are drawn from 
the same parent population cannot be rejected with a probability of 71.9\%.
The close similarity between the neon line ratios of AGN hosts and small-area 
SINGS extractions probably arises from the fact that both probe regions of 
higher ionization parameter.  For the nearby SINGS galaxies, the slit width of 
the short-high module (4\farcs7) only covers the central several hundred 
parsecs, thereby isolating the nuclear region, which most likely has higher 
than average ionization parameter, at the expense of the low-excitation gas 
from the outer regions.  

\begin{figure*}[t]
\centering
\includegraphics[width=0.49\textwidth]{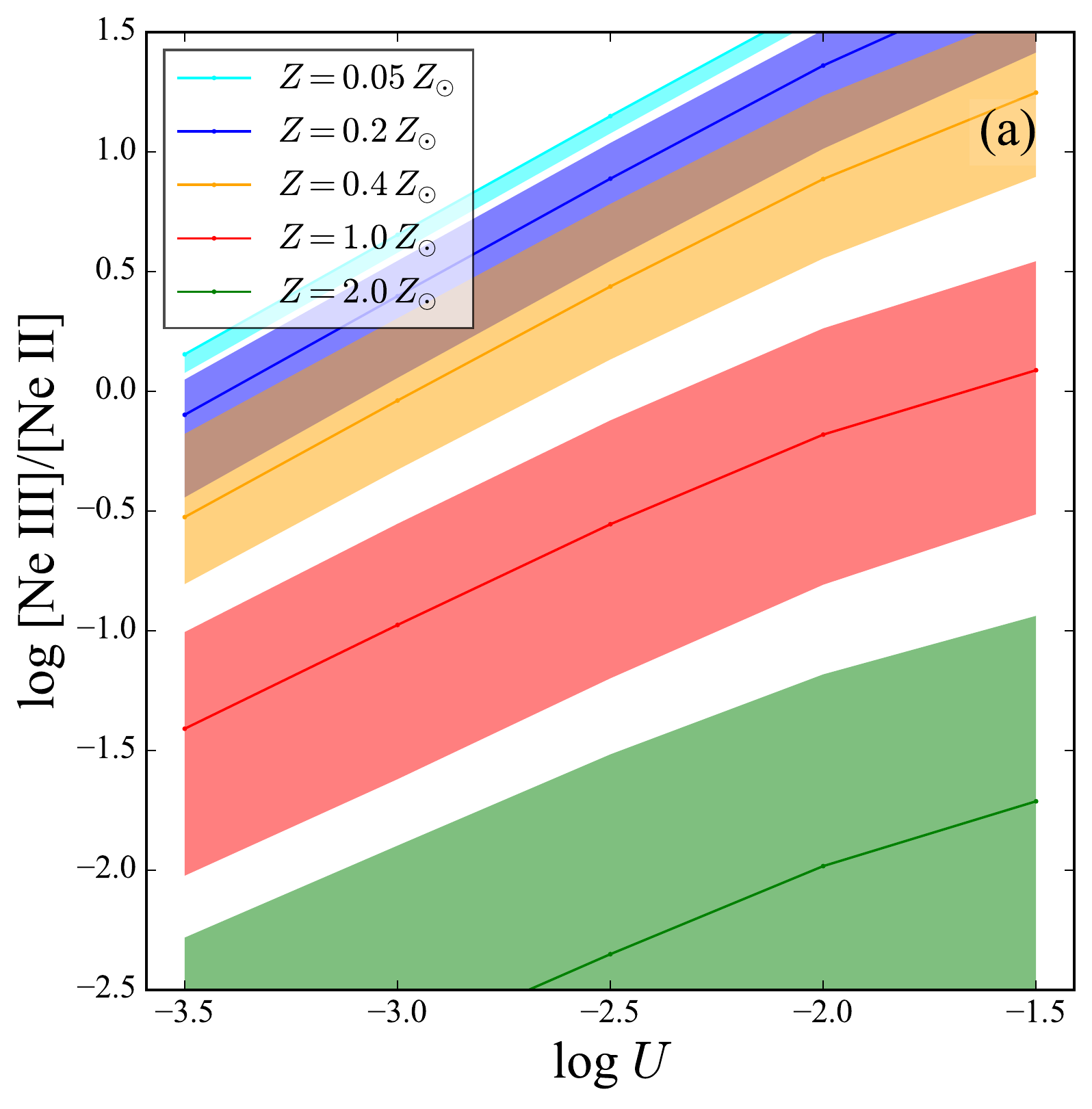}
\includegraphics[width=0.483\textwidth]{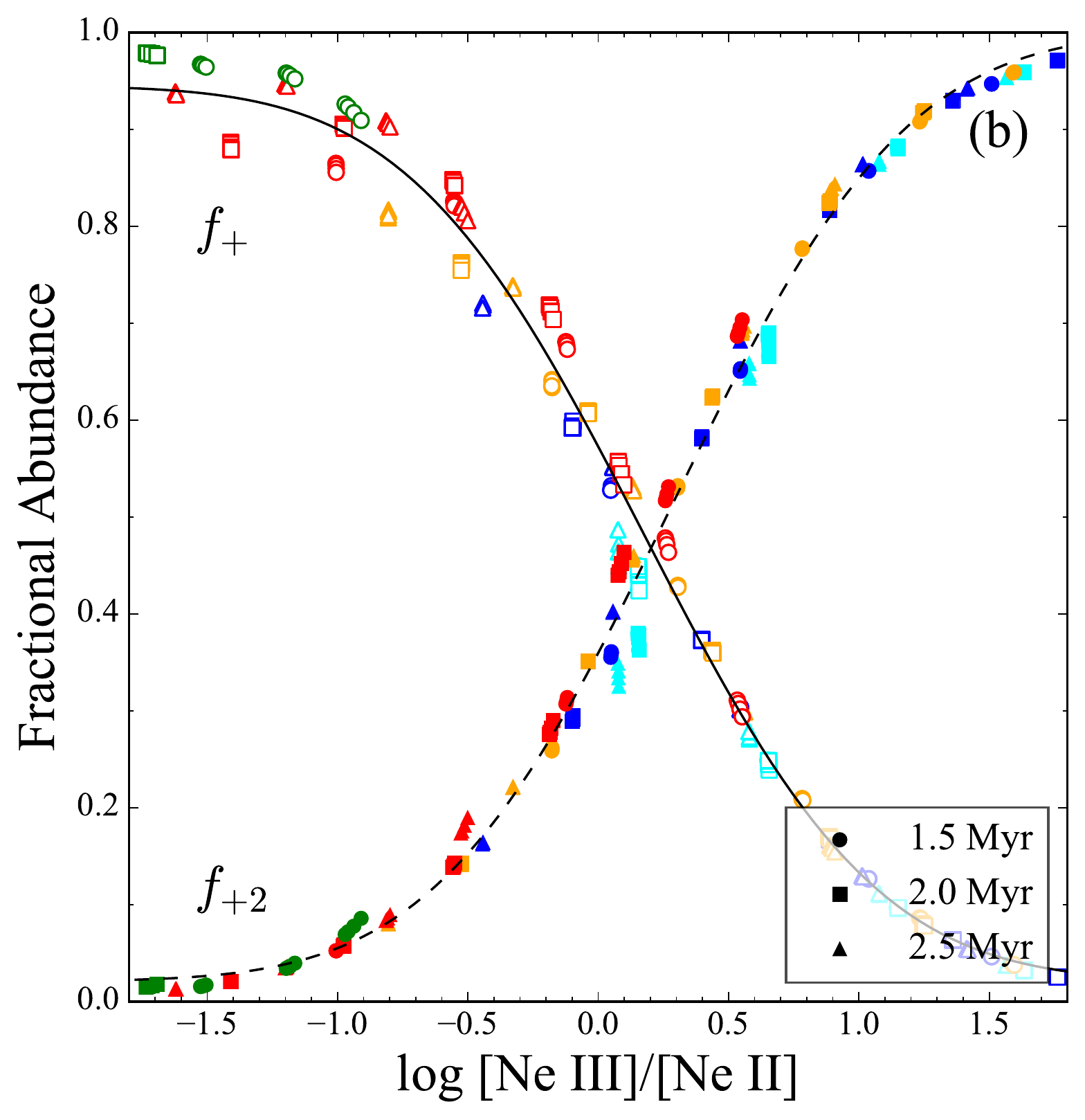}
\caption{(a) The relation between \NeIII/\NeII\ ratio and ionization parameter
$U$ for a starburst with an age of 2 Myr and $n_\mathrm{H} = 10^{2}$ cm$^{-3}$. 
The upper and lower boundaries of the shaded area denote ages of 1.5 and 2.5 
Myr, respectively. Different colors indicate different metallicities. Different
values of $n_\mathrm{H}$ produce almost identical curves, and hence are not 
shown for clarity.  (b) The relation between fractional abundances of \NeII\ 
($f_+$; open symbols) and \NeIII\ ($f_{+2}$; filled symbols) and \NeIII/\NeII\ 
ratio. Colors are the same as in panel (a), showing different metallicities. 
Circles, squares, and triangles represent starbursts with an age of 1.5, 2, and 
2.5 Myr, respectively.  The solid and dashed lines give the best-fit curves for 
$f_+$ and $f_{+2}$ using Equation \ref{eq5}.}
\label{fig4}
\end{figure*}

As discussed in Section \ref{fion}, three parameters affect the \NeIII/\NeII\
ratio in a starburst system: metallicity, ionization parameter, and starburst 
age. Low metallicity, high ionization parameter, and younger starburst age 
enhance \NeIII/\NeII.  High values of \NeIII/\NeII\ are observed in starbursts 
\citep[e.g.,][]{2006ApJ...653.1129B} and blue compact dwarfs \citep[e.g.,][]
{2010ApJ...712..164H}. The \NeIII/\NeII\ ratio in blue compact dwarfs generally
exceeds 1, with extremes up to 18 \citep{2011ApJ...728...45L}. Like blue compact
dwarfs, objects from the SINGS sample with \NeIII/\NeII\ $>$ 2 also have 
sub-solar metallicity.  However, the SINGS objects are generally not 
metal-poor; most have oxygen abundance that are at least solar [12+log(O/H) 
$\gtrsim 8.69$] \citep{2001ApJ...556L..63A}, indicating that high ionization 
parameter, not low metallicity, is responsible for their elevated \NeIII/\NeII.
As the metallicity of the NLR in most AGNs is $\sim 1-2$ times solar (\citealp
{2004ApJS..153....9G, 2004ApJS..153...75G}; but see \citealp{2012ApJ...756...51L}), 
we conclude that the star-forming regions in AGN host galaxies are also 
characterized by high ionization parameters, presumably because they are 
similarly centrally concentrated.

The centrally concentrated star formation we infer from the large \NeIII/\NeII\
ratios of AGN hosts is also supported by observations of PAHs in these systems.
While commonly observed in star-forming galaxies, PAHs are not expected to 
survive the harsh radiation field of an AGN. Yet, PAHs are detected in a 
significant fraction of nuclear regions of Seyfert galaxies, pointing to the 
prevalence of compact central star formation in nearby active galaxies 
\citep{2004ApJ...617..214I}.  \citet{2014ApJ...780...86E} and 
\citet{2018A&A...609A...9L} arrive at similar conclusions based on other 
evidence.

Our inference that star formation in AGN host galaxies occurs preferentially 
in their nuclear regions resonates with the notion that black hole accretion 
and star formation ultimately rely on a common source of gas supply, plausibly
driven to the center of the galaxy by internal dynamical instabilities 
\citep[e.g., a bar:][]{1979MNRAS.187..101L, 2002A&A...392...83B} or
galaxy-galaxy interactions and mergers \citep{2005ApJ...620L..79S, 
2005ApJ...625L..71H, 2015ApJ...804...34H}.

\section{Revisiting the Neon SFR Estimator} \label{revisit}

\citet{2007ApJ...658..314H} first proposed a calibration for SFR based on 
the strength of the low-ionization mid-IR neon lines:

\begin{equation} \label{eq4}
\mathrm{SFR}\, (M_{\odot}\, \mathrm{yr}^{-1}) = 4.34 \times 10^{-41} \left( \frac{L_{\rm \NeII + \NeIII}}{f_+ + 1.67 f_{+2}}\right),
\end{equation}

\noindent 
where $L_{\rm \NeII + \NeIII}$ (in $\mathrm{erg\ s}^{-1}$) is the luminosity of 
\NeII\ 12.81 \micron\ and \NeIII\ 15.56 \micron, and  $f_+$ and $f_{+2}$ are 
the fraction of Ne that is singly and doubly ionized, respectively.  

\subsection{Ionization Fractions for Neon}\label{fion}

\citet{2007ApJ...658..314H} did not consider the influence of neon abundance 
and did not explicitly calculate $f_+$ and $f_{+2}$, which, in principle, can 
vary from galaxy to galaxy, depending, for instance, on the metallicity and 
shape of the ionizing spectrum.  To determine the neon ionization fractions, 
we run simple photoionization models of \HII\ regions with input SEDs from 
{\tt Starburst99} \citep{1999ApJS..123....3L} for instantaneous star formation 
with age 1.5, 2, and 2.5 Myr, a \citet{1955ApJ...121..161S} initial mass 
function with a power-law index $-2.35$, and a lower and upper mass cutoff of 
1 and 100 $M_{\odot}$, respectively.  Starburst ages of 1--3 Myr have previously 
been inferred to reproduce the strong forbidden lines of of Galactic and 
extragalactic \HII\ regions \citep[e.g.,][and references therein]{2011MNRAS.415.3616D}.  
We employ stellar evolutionary tracks with 
high mass loss from the Geneva group \citep{1994A&A...287..803M} 
with $Z=0.05$, 0.2, 0.4, 1.0, and 2.0 $Z_{\odot}$, where $Z_{\odot} = 
0.020$. Other parameters (e.g., wind model and stellar atmosphere) are
kept at their default values, as described in \citet{1999ApJS..123....3L, 
2014ApJS..212...14L}. We assume a 
closed geometry with spherical distribution.  We adopt typical \HII\ region 
\citep[mean of the Orion nebula;][]{2017RMxAA..53..385F} dust grain size 
distribution and consider solar metal abundance with five metallicities ($Z= 
0.05$, 0.2, 0.4, 1.0, and 2.0 $Z_{\odot}$, in order to match the stellar 
metallicity), with dust appropriately scaled with metallicity.  PAHs, 
with properties as in \citet{2008ApJ...686.1125A}, are also included.
We vary the ionization parameter from log $U = -3.5$ to $-1.5$ in steps of 0.5 
dex, and the hydrogen density from log ($n_\mathrm{H}$/cm$^{-3}$) = 1 to 2.5 
in steps of 0.5 dex, covering the typical ranges of star-forming galaxies 
\citep{2013MNRAS.430.2605Z}. The calculation stops when the temperature drops 
below 4000 K (i.e. when reaching the ionization front and almost capturing all 
the \NeII\ and \NeIII). 

Figure \ref{fig4}(a) shows the effect of ionization parameter, metallicity,
and starburst age on the ratio \NeIII/\NeII\ for a 2 Myr starburst. The
\NeIII/\NeII\ ratio is an excellent diagnostic of ionization parameter once the
metallicity and starburst age are fixed, with almost no dependence on 
$n_\mathrm{H}$ \citep[see also][]{2012ApJ...757..108Y}. As metallicity increases, 
the ionizing spectrum becomes softer and more dust contributes to the cooling, 
resulting in a decline in \NeIII/\NeII.  The same trend is seen for starburst 
ages of 1.5 and 2.5 Myr, but the lines are shifted along the vertical axis.  
As the starburst evolves, massive stars die and the rest of the older stellar 
population produces a softer spectrum compared to that of a younger starburst. 
Lower metallicity, younger starburst age, and higher radiation field strength 
all raise the \NeIII/\NeII\ ratio. These parameters can be disentangled using 
other emission-line diagnostics, such as the O32 ratio\footnote{O32 $\equiv$ 
\OIII\ $\lambda5007$/\OII\ $\lambda\lambda3726,\, 3729$} for ionization 
parameter and the $R_{23}$ ratio\footnote{$R_{23} \equiv$ (\OII\ 
$\lambda\lambda 3726,\, 3729$ + \OIII\ $\lambda\lambda 4959, \, 
5007$)/H$\beta$} for metallicity \citep[e.g.,][]{2002ApJS..142...35K, 
2004ApJ...617..240K}, but this is beyond 
the scope of this paper. Although the \NeIII/\NeII\ ratio depends on many 
parameters, we find no dependence of $f_+$ or $f_{+2}$ on parameters other 
than \NeIII/\NeII.  Figure \ref{fig4}(b) shows the relation between fractional 
abundance of \NeII\ ($f_+$) and \NeIII\ ($f_{+2}$) on the \NeIII/\NeII\ ratio. 
The values for different metallicities, ionization parameters, and ages form 
a continuous sequence. The observable \NeIII/\NeII, containing all the 
information on the ionization source and environment, is a good proxy 
of $f_+$ and $f_{+2}$ in star-forming galaxies. 

We use the Levenberg-Marquardt algorithm to fit the relation for all three 
starburst ages together, in the region $-1.5$ $\le$ log \NeIII/\NeII\ $\le$ 
1.5, which covers most of the galaxies.  We obtain

\begin{equation} \label{eq5}
Y = a \times \mathrm{erf} (\log \frac{\rm \NeIII\ }{\rm \NeII\ } + b) + c,
\end{equation}

\noindent
where  $Y$ is $f_+$ or $f_{+2}$, and the error function is 
$\mathrm{erf}(x)=\frac{1}{\sqrt{\pi}} \int_{-x}^{x}e^{-t^2} \mathrm{d}t$.
The best-fit parameters for $f_+$ and $f_{+2}$ are given in Table \ref{table1}. 

\begin{deluxetable}{LRCC}[t]
\tablecaption{Best-fit Parameters for Equation \ref{eq4}
\label{table1}}
\tablehead{
\colhead{$Y$} & \colhead{$a$}& \colhead{$b$}& \colhead{$c$}
}
\startdata
$f_+$ & $-0.462 $ & $-0.174$ & $0.483$\\
$f_{+2}$ & $0.490$ & $-0.281$ & $0.511$\\
\enddata
\end{deluxetable}

\subsection{New SFR Calibrations}

We modify the neon-based SFR calibration as follows.  To include the effect of 
neon abundance, Equation \ref{eq4} should be multipled by $Z_{\odot}/Z$, 
where $Z$ is the metallicity, which scales linearly with the abundance of Ne.  
For star-forming galaxies, Equation \ref{eq4} becomes

\begin{equation} \label{eq6}
\mathrm{SFR}\, (M_{\odot}\, \mathrm{yr}^{-1}) = 4.34 \times 10^{-41} 
(Z_{\odot}/Z) 
\left(\frac{L_{\rm \NeII + \NeIII}}{f_+ + 1.67 f_{+2}}\right),
\end{equation}

\noindent
with $Z$ specified as the user deems appropriate (e.g., from the 
mass-metallicity relation or other more direct information on 
metallicity, when available).  The neon ionization fractions ($f_+$ and 
$f_{+2}$) are given by Equation \ref{eq5}.  In the case of AGNs, 

\begin{equation} \label{eq7}
\begin{split}
\mathrm{SFR}\, (M_{\odot}\, \mathrm{yr}^{-1}) = &4.34 \times 10^{-41} (Z_{\odot}/Z) \\
&\left(\frac{L_{\rm \NeII + \NeIII} - 0.987 L_{\rm \NeV}}{f_+ + 1.67 f_{+2}}\right),
\end{split}
\end{equation}

\noindent
where $L_{\rm \NeV}$ (in $\mathrm{erg\ s}^{-1}$) is the luminosity of \NeV\ 
14.32 \micron, and Equation \ref{eq5} is modified to

\begin{equation} \label{eq8}
Y = a \times \mathrm{erf} (\log \frac{\rm \NeIII - 0.628 \NeV}
{\rm \NeII - 0.345 \NeV} + b) + c.
\end{equation}

\subsection{Comparing Neon with Other SFR Indicators}

\begin{figure*}[t]
\centering
\includegraphics[width=\textwidth]{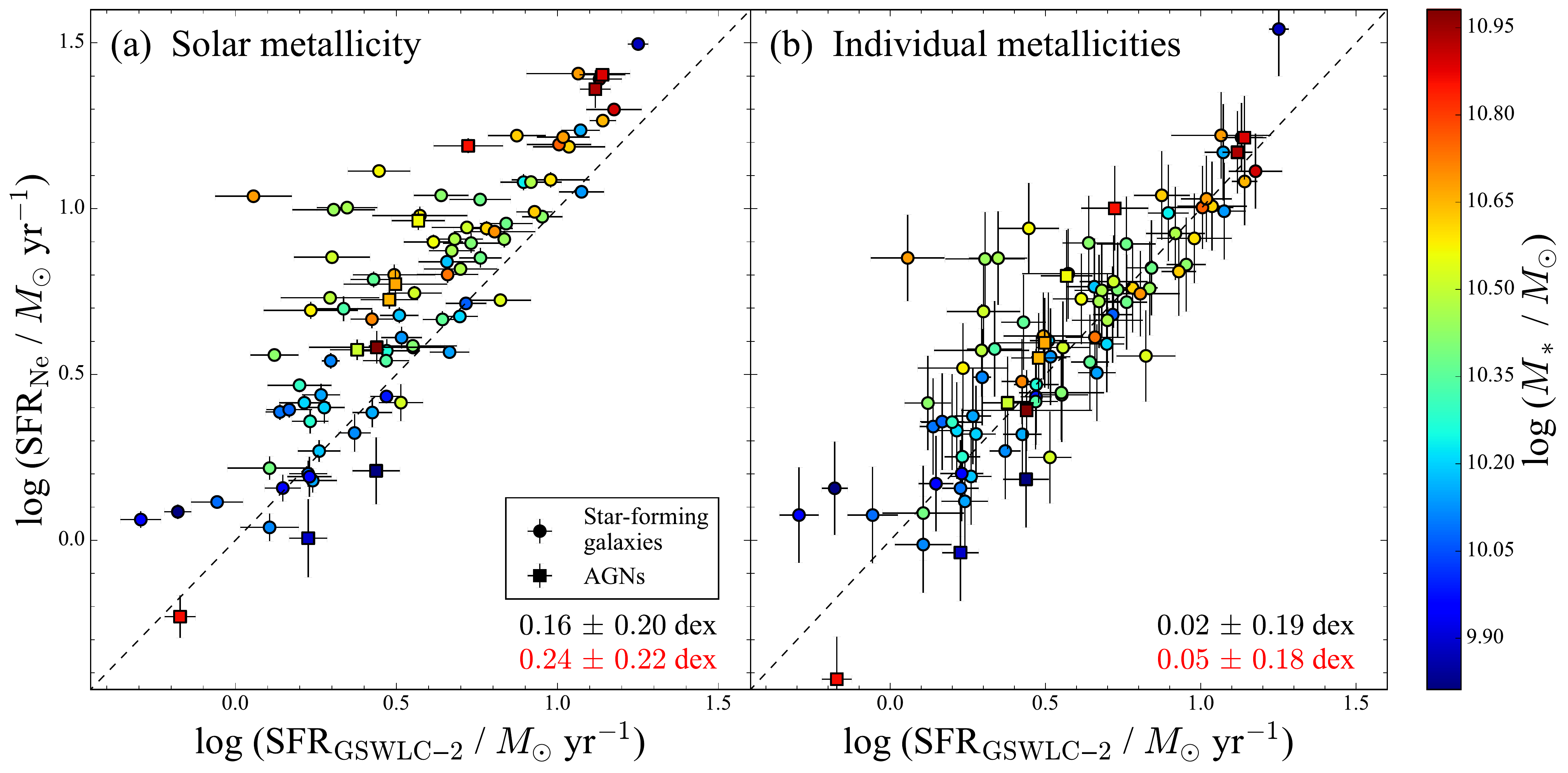}
\caption{Comparison between SFRs derived from SED fitting from GSWLC-2 
\citep{2018ApJ...859...11S} with those from our neon method assuming (a) solar 
metallicity and (b) individual metallicities.  Circles and squares represent 
star-forming galaxies and AGNs, respectively.  Errorbars indicate lower and 
upper 1 $\sigma$ uncertainties. In panel (b), the SFRs from the neon method are
computed assuming individual metallicities derived from the mass-metallicity 
relation.  We adopt the mean of two metallicity estimates (from the N2O2 and 
N2 methods), and the vertical errorbars show the difference between the two 
metallicity calibrations. The data points are color-coded with $M_*$ from 
GSWLC-2, according to the color bar on the right.  The median and standard 
deviation of the difference in SFRs ($\log {\rm SFR}_{\rm Ne} - 
\log {\rm SFR}_{\rm GSWLC-2}$) are shown in the lower-right corner of each 
panel for star-forming galaxies (black) and AGNs (red).}
\label{fig5}
\end{figure*}

We use data for 11/33 AGNs and 78/79 star-forming galaxies from 
\citet{2012ApJ...758....1L} to compare SFRs calculated from our revised method 
with those derived independently to test the consistency between them. We take 
reference SFRs from the second version of the GALEX-SDSS-WISE Legacy Catalog 
\citep[GSWLC-2;][]{2018ApJ...859...11S},  
which provides physical properties (stellar masses, dust attenuations, and 
SFRs) for $\sim$700,000 galaxies with Sloan Digital Sky Survey (SDSS) redshifts
below 0.3 \citep{2016ApJS..227....2S}. GSWLC-2 constrains UV/optical SED fits 
with IR luminosity, obtained from luminosity-dependent IR templates and 
parameterized attenuation curves determined with the Code Investigating GALaxy 
Emission \citep[CIGALE;][]{2018arXiv181103094B} for fitting SEDs.  For AGNs, 
they applied a correction to the IR luminosities to account for dust heating by
the central source.  Following GSWLC-2, we assume a \citet{2003PASP..115..763C} 
initial mass function  and WMAP7 flat cosmology ($H_0=70.4$ km~s$^{-1}$ 
Mpc$^{-1}$, $\Omega_{m}=0.272$).  To obtain consistent metallicities
for both active and inactive galaxies, the metallicities are estimated via the 
mass-metallicity relation \citep{2004ApJ...613..898T, 2008ApJ...681.1183K}
using stellar masses ($M_*$) from GSWLC-2.  Depending on the choice of 
metallicity calibration, absolute metallicities can carry significant uncertainties 
\citep[up to $\sim0.7$ dex;][]{2008ApJ...681.1183K}.  To minimize this 
source of uncertainty, we use the best-fit mass-metallicity relation from 
\citet{2008ApJ...681.1183K} with metallicities derived from a combination of 
the N2O2 method \citep[the ratio of \NII\ $\lambda6584$ to \OII\ 
$\lambda3727$;][]{2002ApJS..142...35K} and the N2 method \citep[the ratio of 
\NII\ $\lambda6584$ to H$\alpha$;][]{2004MNRAS.348L..59P}.  Both of the methods
have low residual discrepancies in relative metallicities and suffer less from 
AGN contamination; the N2O2 method is also less sensitive to ionization 
parameter \citep{2008ApJ...681.1183K}.  The N2O2 method has a high absolute 
metallicity calibration, while the N2 method has low absolute metallicity 
calibration. We adopt the average of the two values as the final metallicity,
with those of N2O2 and N2 as upper and lower boundaries, respectively.

We compare the SFRs derived using our neon method and those from SED fitting.  
With the metallicity fixed to solar (Figure \ref{fig5}a), the neon-based 
SFRs are slightly but systematically higher than those from GSWLC-2, for both 
star-forming and AGN host galaxies, with a clear tendency for the magnitude 
of the excess to correlate with the stellar mass of the system (see color bar).
The offset is marginally higher for AGNs ($0.24\pm0.22$ dex) than for 
star-forming galaxies ($0.16\pm0.20$ dex), because AGN hosts tend to be more 
massive \citep{2003ApJ...583..159H, 2003MNRAS.346.1055K}, and hence more 
metal-rich. After adopting individual metallicities (Figure \ref{fig5}b), the neon-based 
SFRs are shifted downward and come into much better agreement with the 
SED-based SFRs.  The systematic offset between the two methods essentially 
disappears, and the scatter also is reduced somewhat (star-forming galaxies: 
$0.02\pm0.19$ dex; AGNs: $0.05\pm0.18$ dex). We also run models with twice solar 
metallicity.  Although slight differences exist for \NeII/\NeV\ and \NeIII/\NeV\ ratios, 
the resulting SFRs for AGNs are only affected by 0.04 dex (0.09 dex), much 
smaller than the scatter. Furthermore, the overall 
consistency between the two subgroups emphasizes the robustness of our 
procedure for subtracting the NLR contribution to the \NeII\ and \NeIII\ lines 
(Section \ref{agn_contribution}).

\subsection{Future Applications}

The spectroscopic capabilities of {\it Spitzer}/IRS have enabled 
detailed mid-IR diagnostics of the physical conditions in a wide range of 
extragalactic systems, from dwarf galaxies to starbursts and AGNs \citep[e.g.,][]
{2006ApJ...639..157W, 2006ApJ...653.1129B, 2006AJ....132..401B, 2007ApJ...667..149F, 
2010ApJ...725.2270P, 2013ApJ...777..156I}.   The neon-based SFR indicator 
proposed in this work presents a new tool with which to further exploit the 
rich heritage of spectroscopic data contained in the {\it Spitzer}\ archive.
Our technique will also find broad applicability to future infrared facilities, 
which can push investigations of extragalactic star formation to finer 
spatial scales in nearby galaxies and extend spatially integrated studies to 
much higher redshifts.  The mid-infrared instrument 
\citep[MIRI;][]{2015PASP..127..584R} onboard the upcoming {\it James Webb 
Space Telescope} \citep[{\it JWST};][]{2006SSRv..123..485G} will provide
medium-resolution ($R \approx 1500-3500$) integral-field spectroscopy from 4.9 
to 28.8 \micron\ with diffraction-limited spatial resolution longward of 8 
\micron.  This can enable, for example, studies of circumnuclear star formation
out to $z \approx 0.8$.  On a longer timescale, the superior sensitivity of 
the {\it Space Infrared Telescope for Cosmology and Astrophysics} 
\citep[{\it SPICA};][]{2009ExA....23..193S, 2018PASA...35...48E, 
2018PASA...35...30R, 2018PASA...35....2V} promises even greater 
reach to the early Universe.  For instance, the instantaneous wavelength 
coverage of 34--230 \micron\ by the {\it SPICA} Far Infrared Instrument 
(SAFARI) in principle can detect the rest-frame mid-IR neon lines in the
first-generation galaxies and AGNs.

\section{Conclusions}  \label{sec:Conclusions}

We use the photoionization code {\tt CLOUDY} to model the NLR of AGNs, 
incorporating realistic AGN SEDs and input assumptions, with the aim 
of predicting the full, realistic range of intensities for the mid-IR 
fine-structure lines of \NeII\ 12.81 \micron, \NeIII\ 15.56 \micron, and \NeV\ 
14.32 \micron.  Our main conclusions are as follows:

\begin{itemize}

\item We demonstrate that AGNs, over a wide range of realistic conditions, 
emit a relatively restricted range of \NeII/\NeV\ and \NeIII/\NeV\ ratios.  

\item This implies that once \NeV\ is measured, we can estimate the amount of 
\NeII\ and \NeIII\ produced by the AGN, and hence any excess \NeII\ and 
\NeIII\ that can be attributed to star formation from the host galaxy using 
the neon-based SFR estimation method of \citet{2007ApJ...658..314H}.

\item We compute photoionization models of \HII\ regions to obtain an empirical 
relation that enables us to estimate the fractional abundance of singly and 
doubly ionized neon, thereby improving the neon-based SFR estimator for 
galaxies. 

\item Applying our methodology to a sample of low-redshift active galaxies, we 
find that star-forming regions in AGN host galaxies tend to have large 
\NeIII/\NeII\ ratios indicative of high ionization parameters, plausibly a 
consequence of their nuclear environments.  

\item We update the neon-based SFR estimator by explicitly including 
metallicity as a key factor.

\item The good consistency between the neon-based SFRs and those independently 
derived from SED fitting emphasizes the effectiveness of our method for 
correcting for NLR contamination and the overall robustness of the newly 
proposed SFR calibration for AGNs.

\item 
Our method provides a promising tool to explore star formation in 
active and inactive galaxies across cosmic time with future IR facilities such 
as {\it JWST} and {\it SPICA}.
\end{itemize}

\acknowledgments
We thank an anonymous referee for helpful comments and suggestions. 
This work was supported by the National Key R\&D Program of China 
(2016YFA0400702) and the National Science Foundation of China (11721303).

\end{document}